\documentclass[11pt]{article}
\usepackage{pst-all}
\usepackage{pstricks}
\usepackage{pstcol,pst-fill,pst-grad}
\usepackage{amsfonts}
\usepackage{graphicx}
\usepackage{amsmath}

\makeatletter \@addtoreset{equation}{section}

\newcommand{\be}{\begin{equation}}
\newcommand{\ee}{\end{equation}}
\newcommand{\bea}{\begin{eqnarray}}
\newcommand{\eea}{\end{eqnarray}}
\begin{document}
\date{}
\title{
\textbf{    Four-qubit Systems  and   Dyonic  Black Hole-Black Branes  in   Superstring Theory  }\\
\textbf{   } }
\author{ A. Belhaj$^{1,2}$, M. Bensed$^{3}$,  Z. Benslimane$^{3}$,  M.  B. Sedra$^{3}$, A. Segui$^{2}$
\hspace*{-8pt} \\
 {\small $^{1} $LIRST, D\'epartement de
Physique,   Facult\'e Polydisciplinaire, Universit\'e Sultan Moulay
Slimane}\\{ \small B\'eni Mellal, Morocco }
\\ {\small $^{2}$    Departamento de F\'{i}sica T\'e{o}rica, Universidad de Zaragoza, E-50009-Zaragoza, Spain }\\  {\small $^{3}$    D\'{e}partement de
Physique, LabSIMO,  Facult\'{e} des Sciences, Universit\'{e} Ibn
Tofail }\\{ \small K\'{e}nitra, Morocco}}  \maketitle

\begin{abstract}
Using dyonic solutions in the  type IIA superstring theory  on
Calabi-Yau manifolds, we reconsider the study of  black objects  and
quantum information theory using string/string duality in six
dimensions.
 Concretely,  we relate four-qubits with  a stringy quaternionic moduli space
of type IIA compactification associated with a dyonic black solution
formed by black holes (BH) and  black 2-branes (B2B) carrying 8
electric charges and 8 magnetic charges. This connection is made by
associating the cohomology classes of  the heterotic superstring on
$T^{4}$ to four-qubit states. These states are  interpreted  in
terms of   such dyonic  charges resulting from  the quaternionic
symmetric space $\frac{SO(4,4)}{SO(4)\times SO(4)}$  corresponding
to  a $N=4$ sigma model superpotential  in two dimensions. The
superpotential is considered as a functional depending on four
quaternionic fields mapped  to
 a class of  Clifford algebras denoted  as  $Cl_{0,4}$. A  link   between
 such an algebra and
the cohomology classes of $T^4$  in heterotic superstring theory is
also  given.
\end{abstract}

\thispagestyle{empty}

\newpage{}\setcounter{page}{1} \newpage{}

\section{Introduction}

Extremal black branes have been extensively studied in the framework
of superstring theory on  the Calabi-Yau (CY)  manifolds
\cite{1,2,3,4}. These black solutions have been approached by
exploring the attractor mechanism and  the topological string theory
\cite{5,6,7,8,9}. In the  attractor mechanism scenario, the scalars
could  be fixed in terms of the black brane charges by extremising
the associated  potential with respect to the stringy  moduli
obtained from the superstring  theory  compactified  on the
Calabi-Yau manifolds. Moreover, the corresponding entropy functions
have been computed using the string duality symmetries acting on the
invariant black brane charges. In this way, several Calabi-Yau
compactifications have been examined producing various results
dealing with black objects in type II superstrings using D-brane
physics \cite{10,11}.

The black  objects, embedded in  superstring theory
compactifications,  can be  connected to quantum information theory
 using the qubit
analysis {[}12-24{]}.  Precisely, a fascinating correspondence has
been discovered between  quantum information theory  and superstring
theory. The main obtained  relations  are  between the entropy
formulas for specific black hole solutions in supergravity theories
and entanglement measures for  certain multiqubit systems
\cite{17,18}. Alternative studies have been conducted using toric
geometry and graph theory \cite{24,25,250,251}. The underlying idea
is a link between the $N=2$ STU black hole charges and three-qubit
states which has been established in \cite{15,16}. Furthermore,  the
analysis based on three-qubits has been developed to describe the
structure of extremal black hole solutions  in terms of four-qubit
systems\cite{12,19,22,23,26}.  In all the works on  four-qubits, one
uses the
 complex geometry to deal with the corresponding  black hole entropy.   For more details,   we refer to
\cite{22,260}.

The  main goal of this work  is to contribute to these activities by
approaching   four-qubit systems   using  string   dualities and a
quaternionic description of stringy moduli spaces. Concretely, we
reconsider the investigation of black objects and quantum
information theory using string/string duality  in the context of
dyonic solutions in type II superstring compactifications on
Calabi-Yau manifolds. This  may  offer a new take on the moduli
space of black objects and quantum information theory.
 More precisely,  we link four-qubits with  a stringy quaternionic moduli space
of type IIA  superstring compactification  associated with a dyonic
black solution formed by black holes (BH) and back 2-branes (B2B),
referred to as $\left(\begin{array}{c}
0\\
2
\end{array}\right)$ dyonic object with eight electric charges and  eight magnetic charges, producing sixteen charges. This connection is made
 by associating  the cohomology
classes of the heterotic superstring on $T^{4}$ to four-qubit
states. These states can  be interpreted in terms of charges of such
dyonic solutions resulting from the  quaternionic symmetric space
$\frac{SO(4,4)}{SO(4)\times SO(4)}$  corresponding to a
superpotential of $N=4$ sigma model  in two dimensions. The
superpotential has been considered  as  a functional depending on
quaternionic fields related to a class of the Clifford algebras
 developed in \cite{27}. This  algebra,   denoted  as  $Cl_{0,4}$, provides  a link  with  the cohomology classes of
$T^4$ in the heterotic superstring  compactification.

The organization of the paper is as follows. Section 2 is a concise
review  on the study of dyonic solutions in type II superstrings
compactified on  $n$-dimensional Calabi-Yau manifolds. We emphasize
the black solutions carrying charges in $10-2n$ dimensions sharing
electric and magnetic  charge dualities. A classification according
to black brane dimensions results in two dyonic solutions which can
 be generalized in arbitrary dimension.  Section 3 contains a  link
 between
four-qubit systems and    a  quaternionic geometry   considered  as
a reduction of  the moduli space of  the heterotic superstring
compactified on $T^4$ being dual to type IIA superstring on the K3
surface. This  moduli space will be considered as the moduli space
of a dyonic BH-B2B object with 8 electric charges and 8 magnetic
charges. In section 4, the four-qubit states are related to dyonic
charges living in the  moduli space $\frac{SO(4,4)}{SO(4)\times
SO(4)}$.  In section 5, we claim that the usual decomposition of
$SO(4)\times SO(4)\longrightarrow SU(2)\times SU(2)\times
SU(2)\times SU(2)$ results in four quaternionic fields that can be
interpreted  in terms of  four-qubit states. We suggest that these
fields  can produce a quaternionic superpotential associated with a
$N=4$ sigma model in two dimensions \cite{270}. Precisely, we regard
the superpotentiel as an element of a particular class of the
Clifford algebras denoted as  $Cl_{0,4}$.  We end in Section 6 with
some
 discussions and open
questions.

\section{Dyonic solutions in type II superstrings }
We start by reconsidering the study  of dyonic solutions in type II
superstrings compactified on CY manifolds, that we will later need.
It is recalled that a $n$-dimensional CY manifold (CY $n$-folds) is
a complex with the K\"{a}hler structure. The latter involves a
global nonvanishing holomorphic $n$-form which is equivalently to a
K\"{a}hler manifold with a vanishing first Chern class $c_{1}=0$
required by a $SU(n)$ holonolmy group \cite{28}. The superstring
compactification on such manifolds preserves only
$\frac{1}{2^{n-1}}$ of the  ten dimensional supercharges. It has
been remarked that each manifold is associated with a Hodge diagram
playing an important r\^ole in the determination of the superstring
theory spectrum in $10-2n$ dimensions \cite{28,29,30,31}. To have a
general idea on such data, we list the Hodge diagrams of the $T^{2}$
torus, the K3 surface and the CY threefolds, respectively:
\[
\begin{tabular}{|l|l|l|}
\hline  \ensuremath{n=1}  &  \ensuremath{\begin{tabular}{lll}
  &  \ensuremath{h^{0,0}}  &  \\
\ensuremath{h^{1,0}}  &   &  \ensuremath{h^{0,1}} \\
 &  \ensuremath{h^{1,1}}  &
\end{tabular}}  &  \ensuremath{\begin{tabular}{lll}
  &  \ensuremath{1}  &  \\
\ensuremath{1}  &   &  \ensuremath{1} \\
 &  \ensuremath{1}  &
\end{tabular}} \\
\hline \ensuremath{n=2}  &  \ensuremath{\begin{tabular}{lllll}
  &   &  \ensuremath{h^{0,0}}  &   &  \\
 &  \ensuremath{h^{1,0}}  &   &  \ensuremath{h^{0,1}}  &  \\
\ensuremath{h^{2,0}}  &   &  \ensuremath{h^{1,1}}  &   &  \ensuremath{h^{0,2}} \\
 &  \ensuremath{h^{2,1}}  &   &  \ensuremath{h^{1,2}}  &  \\
 &   &  \ensuremath{h^{2,2}}  &   &
\end{tabular}}  &  \ensuremath{\begin{tabular}{lllll}
  &   &  \ensuremath{1}  &   &  \\
 &  \ensuremath{0}  &   &  \ensuremath{0}  &  \\
\ensuremath{1}  &   &  \ensuremath{20}  &   &  \ensuremath{1} \\
 &  \ensuremath{0}  &   &  \ensuremath{0}  &  \\
 &   &  \ensuremath{1}  &   &
\end{tabular}} \\
\hline \ensuremath{n=3}  &  \ensuremath{\begin{tabular}{lllllll}
  &   &   &  \ensuremath{h^{0,0}}  &   &   &  \\
 &   &  \ensuremath{h^{1,0}}  &   &  \ensuremath{h^{0,1}}  &   &  \\
 &  \ensuremath{h^{2,0}}  &   &  \ensuremath{h^{1,1}}  &   &  \ensuremath{h^{0,2}}  &  \\
\ensuremath{h^{3,0}}  &   &  \ensuremath{h^{2,1}}  &   &  \ensuremath{h^{1,2}}  &   &  \ensuremath{h^{0,3}} \\
 &  \ensuremath{h^{3,1}}  &   &  \ensuremath{h^{2,2}}  &   &  \ensuremath{h^{1,3}}  &  \\
 &   &  \ensuremath{h^{3,2}}  &   &  \ensuremath{h^{2,3}}  &   &  \\
 &   &   &  \ensuremath{h^{3,3}}  &   &   &
\end{tabular}}  &  \ensuremath{\begin{tabular}{lllllll}
  &   &   &  \ensuremath{h^{0,0}}  &   &   &  \\
 &   &  \ensuremath{0}  &   &  \ensuremath{0}  &   &  \\
 &  \ensuremath{0}  &   &  \ensuremath{h^{1,1}}  &   &  \ensuremath{0}  &  \\
\ensuremath{1}  &   &  \ensuremath{h^{2,1}}  &   &  \ensuremath{h^{1,2}}  &   &  \ensuremath{1} \\
 &  \ensuremath{0}  &   &  \ensuremath{h^{1,1}}  &   &  \ensuremath{0}  &  \\
 &   &  \ensuremath{0}  &   &  \ensuremath{0}  &   &  \\
 &   &   &  \ensuremath{1}  &   &   &
\end{tabular}}
\\\hline \end{tabular}
\]
\global\long\def\m#1{\makebox[10pt]{\ensuremath{#1}}}

 It turns out that $h^{p,q}$ denotes  the number of the holomorphic and the anti holomorphic $(p,q)$
forms. Deleting the zeros, one observes that each diagram contains
two central orthogonal lines. For the CY $n$-folds, the vertical
line encodes the parameters describing  the K\"{a}hler deformations.
It has been shown  that the number of such size parameters,
representing the K\"{a}hler deformations of the metric, is fixed by
$h^{1,1}$. The horizontal one represents the parameters of the
complex structure (shape parameters) given by $h^{n-1,1}$. Beside
these parameters, the Hodge diagram can be explored to produce the
all physical data in lower dimensional superstring theory
compactifications. Indeed, the moduli space of the CY type II
superstring compactifications is determined by the geometric
deformations of the CY metric including the antisymmetric B-field of
the \textbf{NS-NS} sector, the dilaton, specifying the string
coupling constant and the scalars derived from  the \textbf{R-R}
gauge fields on non trivial cycles of the CY spaces. In connection
with the  black solutions in the type II superstring
compactifications, these scalar fields which are associated with
supergravity models having  $2^{6-n}$ supercharges are coupled to an
abelian gauge symmetry providing electric and magnetic charges of
black objects in $10-2n$ dimensions. In this way, the near horizon
of these black objects is usually defined by the product of Ads
spaces and spheres as follows
\begin{equation}
Ads_{p+2}\times S^{8-2n-p},
\end{equation}
where $p$ is the internal dimension of the black brane. $n$ and $p$
verify the following constraint
\begin{equation}
2\leq8-2n-p.
\end{equation}
In the compactified theory living in $10-2n$, the electric/magnetic
duality liking a $p$-dimensional electrical black brane to a
$q$-dimensional magnetic one is assured  by the constraint
\begin{equation}
p+q=6-2n.\label{gd}
\end{equation}
A priori,  this equation can be solved in different ways according
to the $(p,q)$ couple values. The solution can be classified as
follows
\begin{itemize}
\item $(p,q)=(0,6-2n)$ describing an electrical charged black hole (BH)
\item $(p,q)=(3-n,3-n)$ describing dyonic black branes (DB$(3-n)$B)
\item $(p,q)\neq(0,6-2n)$ and $(p,q)\neq(3-n,3-n)$, describing black objects
like strings, membranes and higher-dimensional branes.
\end{itemize}
However, a closed inspection shows that we have two kinds of dyonic
solutions carrying electric and magnetic charges by considering
objet like doublets. They are listed as follows
\begin{enumerate}
\item a solution$\left(\begin{array}{c}
3-n\\
3-n
\end{array}\right)$ consisting of the same object associated with
\begin{equation}
p=q=3-n.
\end{equation}

\item a solution $\left(\begin{array}{c}
p\\
6-2n-p
\end{array}\right)$ consisting of an electrically charged black object and its magnetic
dual one  corresponding to
\begin{equation}
p,\quad q=6-2n-p,\qquad p\neq3-n.
\end{equation}

\end{enumerate}
The last solution is considered as a single object sharing similar
features of the usual  dyonic solution described by the same object.
We believe  that one could build it in any dimension. To see that,
let us consider a model obtained by the compactification of type IIA
superstring on the K3 surface required by the
$h^{1,0}=h^{1,2}=h^{0,1}=h^{2,1}=0$. Indeed, it is recalled that the
type IIA superstring perturbative bosonic massless sector contain
the following fields
\begin{equation}
\text{{\bf NS-NS}}:g_{MN},\;\;B_{MN},\;\;\phi\qquad\text{{\bf
R-R}}:A_{M},\;\;C_{MNK}
\end{equation}
where $M,N,K=0,\ldots,9$. This compactification produces a $N=2$
supergravity in six dimensions with the following bosonic spectrum
\begin{equation}
g_{\mu\nu},\;\;B_{\mu\nu},\;\;\phi,\;\;A_{\mu},C_{\mu\nu\rho},\;\;C_{\mu
ij},\quad \phi_{\alpha},\;\alpha=1,\ldots,80.
\end{equation}
In this spectrum, $g_{\mu\nu}$ is the six dimensional graviton
metric, $B_{\mu\nu}$ and $C_{\mu\nu\rho}$ are the six dimensional
antisymmetric gauge fields. The field $A_{\mu}$ and $C_{\mu ij}$
represent the gravi-photon and Maxwell gauge fields in six
dimensions. These fields are obtained from the compactification of
$C_{\mu\nu\rho}$ on the real 2-cycles of the K3 surface. Since
$C_{\mu\nu\rho}$ is dual to a vector in six dimensions, the theory
has an $U(1)^{24}$ abelian gauge symmetry. Besides the antisymmetric
gauge field $B_{\mu\nu}$, the total gauge symmetry reads
\begin{equation}
G=G_{1}\times G_{2}=U(1)^{24}\times U(1)
\end{equation}
associated with one and two-form  gauge fields, respectively. In six
dimensions, these gauge fields are coupled to the scalar fields. In
addition to the dilaton $\phi$ in six dimensions, there are $80$
scalar fields $\phi_{\alpha}$ which can be arranged to form the
moduli space of type IIA superstring on the K3 surface. The latter
can be viewed as a scalar manifold of half-maximal, non-chiral Type
IIA supergravity in six dimensions, coupled to 20 vector multiplets,
which reads as
\begin{equation}
\frac{SO(4,20)}{SO(4)\times SO(20)}\times SO(1,1).\label{modulik3}
\end{equation}
It has been shown that the first factor $\frac{SO(4,20)}{SO(4)\times
SO(20)}$ represents the geometric deformations of the K3 surface in
the presence of the antisymmetric B-field of the \textbf{NS-NS}
sector and it is linked to  the symmetry group $G_{1}=U(1)^{24}$.
However,  the second factor $SO(1,1)$  represents  the dilaton
scalar field  which is associated with $G_{2}=U(1)$ \cite{31,310}.
It has been remarked that the space (\ref{modulik3}) is related to
the electric/magnetic duality assured by the condition
\begin{equation}
p+q=2
\end{equation}
producing two possible dynonic solutions
\begin{enumerate}
\item $\left(\begin{array}{c}
0\\
2
\end{array}\right)$ BH-B2B with near-horizon geometries $Ads_{2}\times S^{4}$ and $Ads_{4}\times S^{2}$
\item  $\left(\begin{array}{c}
1\\
1
\end{array}\right)$ black string (BS) with the near horizon geometry $Ads_{3}\times S^{3}.$
\end{enumerate}
In six dimensions, the factor $\frac{SO(4,20)}{SO(4)\times SO(20)}$
corresponds to the dyonic BH-B2B object $\left(\begin{array}{c}
0\\
2
\end{array}\right)$ with 24 electric charges and 24 magnetic charges. It can be realized
in terms of a D-brane  system containing $\{D0,D2,D4,D6\}$ which can
be  placed on the corresponding Hodge diagram. The 24+24 charges are
represented by the configuration of the K3 surface as follows
\begin{equation}
{\arraycolsep=2pt\begin{array}{ccccc}
 &  & \m 1\\
\\
\m 1 &  & \m{20} &  & \m 1\\
\\
 &  & \m 1
\end{array}}\;\equiv\;{\arraycolsep=2pt\begin{array}{ccccc}
 &  & (D0,D2)\\
\\
(D2,D4) &  & (D2,D4) &  & (D2,D4)\\
\\
 &  & (D4,D6)
\end{array}}
\end{equation}
However,  the factor $SO(1,1)$  is associated with  the dyonic BS
$\left(\begin{array}{c}
1\\
1
\end{array}\right)$ having  one electric charge and one magnetic
charge. It is  represented by the following D-brane  configuration
on the K3 surface
\begin{equation}
{\arraycolsep=2pt\begin{array}{ccccc}
 &  & \m 1\\
\\
\m 1 &  & \m{20} &  & \m 1\\
\\
 &  & \m 1
\end{array}}\;\equiv\;{\arraycolsep=2pt\begin{array}{ccccc}
 &  & F1\\
\\
 & . & . & .\\
\\
 &  & NS5
\end{array}}
\end{equation}
Having discussed the compactification of the  type IIA superstring
on the K3 surface, we would like to relate the corresponding  black
objects and quantum information theory by combining string/string
duality and non trivial cycles  appearing in  the toroidal
compctification of the  heterotic superstring on $T^4$. A special
emphasis is put on four-qubit systems which will be linked with  a
particular quaternionic geometry  considered as a subspace of the
one appearing in (\ref{modulik3}).

\section{Four-qubit systems  and  dyonic solutions on the  symmetric space  $
\frac{SO(4,4)}{SO(4)\times SO(4)}$} Several structural similarities
between  quantum information theory  and superstring theory have
been established forming the so called black hole qubit
correspondence (BHQC). The first  mapping  was between the entropy
formulae of certain black holes and the entanglement measures of
qubit systems using Cayley's hyperdeterminant \cite{32,320,321}. In
particular, it has been shown that the square root of Cayley's
hypertederminant is linked to eight charges of extremal black holes
in the STU model by the entropy formulae as follow
\begin{equation} S = \pi \sqrt{|Det a_{ABC}|} = \frac{\pi}{2}
\sqrt{\tau_{ABC}}.
\end{equation}
Here  $\tau_{ABC}$ and $S$  are the 3-tangle measure and the black
hole entropy,  respectively. Furthermore, it is  clearly interesting
to recall  that the Cayley's hyperderminant denoted  as $Det A$ is
defined as
\begin{equation}
Det A \equiv-\frac{1}{2}\epsilon^{A_{1}A_{3}}\epsilon^{A_{2}A_{4}}
\epsilon^{B_{1}B_{2}}\epsilon^{B_{3}B_{4}}\epsilon^{C_{1}C_{2}}\epsilon^{C_{3}C_{4}}
a_{A_{1}B_{1}C_{1}}a_{A_{2}B_{2}C_{2}}a_{A_{3}B_{3}C_{3}}a_{A_{4}B_{4}C_{4}}.
\end{equation}
It is not hard to see that this is an homogeneous quartic polynomial
that involves an interesting physical interpretation in terms of the
STU black hole charges embedded in type II superstrings.  Precisely,
the solution of  the STU black hole, in the case of spherical
symmetry,  is given in terms of 8 charges $(q_0, q_1 , q_2, q_3,
p^0, p^1, p^2, p^3)$. In this way,  the square of the extremal STU
black hole entropy  is proportional to a quartic polynomial of $q_0,
q_1 , q_2, q_3, p^0, p^1, p^2$ and $p^3$ \cite{15,16}
\begin{equation}
\begin{split}
S^2 = & \quad \pi^2 \{ -(p^0q_0 + p^1q_1 + p^2q_2+ p^3q_3)^2 \\ &+
4((p^1q_1)(p^2q_2)+(p^1q_1)(p^3q_3)+ (p^3q_3)(p^2q_2)
\\ &+q_0p^1p^2p^3 - p^0q_1q_2q_3) \}.
\end{split}
\end{equation}
Under a suitable  mapping, the eight charges of  the STU black hole
correspond to the states of  a three-qubit system as
follows\cite{40}
\begin{center}
$\left(\begin{array}{c}
q_{0}\\
q_{1}\\
q_{2}\\
q_{3}\\
p^{0}\\
p^{1}\\
p^{2}\\
p^{3}
\end{array}\right) \qquad \longleftrightarrow \qquad \left(\begin{array}{c}
a_{000}\\
-a_{001}\\
-a_{010}\\
-a_{100}\\
a_{111}\\
a_{110}\\
a_{101}\\
a_{011}
\end{array}\right).$
\end{center}

Motivated by  the STU black hole,  three-qubits  and certain
extended works, we would like to link the four-qubits with a
particular stringy  moduli space given by
\begin{equation} \label{so44}
 \frac{SO(4,4)}{SO(4)\times SO(4)}.
\end{equation}
 We will show that this  moduli space can be considered as a
subpart of a general quaternionic  geometry, related to the c-map
image of the symmetric projective special Kähler manifold
$SL(2,R)^3/U(1)^3$, which is the vector multiplets scalar manifold
of the $N=2$  $D=4$ STU supergravity model \cite{41}. Then, we will
show that this moduli space corresponds to a dyonic black object in
six dimensions carrying eight  electric and eight magnetic charges
playing the same role as a black string in six dimensions.
Concretely,   these black solutions are mapped to four-qubit
physical systems using string dualities. It is recalled that the
physics of the qubit has been extensively investigated from
different physical and mathematical aspects \cite{33,34,35}. Using
Dirac notation, one-qubit is described by the following state
\begin{equation}
|\psi\rangle=a_{0}|0\rangle+a_{1}|1\rangle.
\end{equation}
Here,  $a_{i}$ are  considered as  complex numbers verifying the
probability condition
\begin{equation}
|a_{0}|^{2}+|a_{1}|^{2}=1.
\end{equation}
It should be denoted that this condition can be interpreted
geometrically in terms of the so-called Bloch sphere,
$\mathbb{CP}^{1}$. Similarly, the two-qubits are represented by the
general state
\begin{equation}
|\psi\rangle=a_{00}|00\rangle+a_{10}|10\rangle+a_{01}|01\rangle+a_{11}|11\rangle.
\end{equation}
In this case, the probability condition is
\begin{equation}
|a_{00}|^{2}+|a_{10}|^{2}+|a_{01}|^{2}+|a_{11}|^{2}=1,
\end{equation}
defining a 3-dimensional complex projective space $\mathbb{CP}^{3}$
generalizing the Bloch sphere. This analysis can be extended to
$N$-qubits  having  $2^{N}$ configuration states. For instance, the
general state  of the  four-qubits reads as
\begin{equation}
|\psi\rangle=\sum\limits
_{ijk\ell=0,1}a_{ijk\ell}|ijk\ell\rangle,\label{qudit}
\end{equation}
where $a_{ijk\ell}$ verify the  normalization condition
\begin{equation}
\sum\limits
_{ijk\ell=0,1}a_{ijk\ell}\overline{a}_{a_{ijk\ell}}=1\label{pcn}
\end{equation}
defining the $\mathbb{CP}^{15}$ complex projective space.  Using
results on  string dualities, these states will be linked to charges
of a dyonic object embedded in type IIA superstring.

Roughly speaking, the  moduli space (\ref{so44}) can be considered
as  a particular  geometry of
\begin{equation}
\frac{SO(4,m)}{SO(4)\times SO(m)}\times SO(1,1)
\end{equation}
where $m\geq3$ is an integer which can be fixed by the
compactification  in question.  It  appears  naturally in six
dimensional supergravity models.

The link that we are after push us to consider the case $m=4$
reducing the above moduli space to
\begin{equation}
\frac{SO(4,4)}{SO(4)\times SO(4)}\times SO(1,1).
\end{equation}
It turns out that in the analysis of the  superstring
compactifications (sigma model fields),  we can remove the factor
$SO(1,1)$ by fixing the  dilaton. The remaining factor can be
obtained using different ways. A possible one is to think about the
decomposition of the moduli space of the K3 surface, or the
heterotic  superstring on $T^{4}$. It has been shown that such two
models are equivalent. This duality is known by string/string
duality in six dimensions \cite{36}. Indeed,  it  is possible to use
the following decomposition
\begin{equation}
\frac{SO(4,20)}{SO(4)\times
SO(20)}\rightarrow\frac{SO(4,4)}{SO(4)\times
SO(4)}\times\frac{SO(4,16)}{SO(4)\times SO(16)}
\end{equation}
supported by the fact that
\begin{equation}
4\times 20=4\times 4+ 4\times 16.
\end{equation}
Examining the string/string duality in six dimensions, the factor
$\frac{SO(4,16)}{SO(4)\times SO(16)}$ corresponds to the twistor
sector in type IIA superstring side. It is associated with the fixed
points of the orbifold compactification. It is  interesting   to
note that this sector has played a primordial r\^ole in solving a
serious problem in type IIA spectrum in six dimensions being the
absence of non abelian gauge symmetries. However,  this sector will
be ignored and we consider only the factor
$\frac{SO(4,4)}{SO(4)\times SO(4)}$. This will be done by
restricting quaternionic dimensions living in six dimensional
supergravity  where certain  parts  of stringy moduli space should
take zero values. In superstring theory, this  factor can be
obtained from the toroidoal compactification of  the heterotic
superstring by ignoring the contribution of  the  gauge symmetry
derived from the 26-dimensional  bosonic sector. In particular, the
compactification of the heterotic superstring on $T^{4}$ produces
$\frac{4\times5}{2}$ degrees of freedom associated with the metric
$g_{ij}$ and $\frac{4\times3}{2}$ degrees of freedom corresponding
to the anti-symmetric field $B_{ij}$. Then, we have $4\times4=16$
real scalars  parameterizing the symmetric space
$\frac{SO(4,4)}{SO(4)\times SO(4)}$. Besides such scalar fields, we
have also the configuration representing  the abelian gauge fields.
These abelian gauge fields can be obtained from the $B_{\mu i}$ and
$g_{\mu i}$ fields. This generates the gauge symmetry
\begin{equation}
G_{1}=U(1)^{4}\times U(1)^{4}
\end{equation}
which provides a $SO(4)\times SO(4)$ isotopy symmetry. In six
dimensions, this symmetry corresponds to
\begin{itemize}
\item 8 electric charges of BH  solution with the $AdS_{2}\times S^{4}$
near horizon geometry.
\item 8 magnetic charges of a B2B  solution with the $AdS_{4}\times S^{2}$
near horizon geometry.
\end{itemize}
Following the general discussion made in the previous section, the
moduli space $\frac{SO(4,4)}{SO(4)\times SO(4)}$ corresponds to a
BH-B2B dyonic  object $\left(\begin{array}{c}
0\\
2
\end{array}\right)$ with 8 electric charges and 8 magnetic charges. It can be realized
in terms of a  D-brane system containing $\{D0,D2,D4,D6\}$ which can
be  placed on the Hodge diagram of $T^{4}$. At the level, it is
intersecting to note that this solution could generate a single
dyonic object in four dimensions. Assuming  that the corresponding
gauge fields  survive in four dimensions, the B2B can be converted
to a BH in four dimensions using   a possible compactification on a
2-sphere $S^{2}$
\begin{equation}
B2B\stackrel{S^{2}}{\longrightarrow}\ BH.
\end{equation}
 This could produce a single dyonic solution$\left(\begin{array}{c}
0\\
0
\end{array}\right)$ in four dimensions with the near-horizon geometry $AdS_{\text{2}}\times S^{2}$
by thinking $AdS_{4}\times S^{2}$ as $AdS_{2}\times S^{2}\times
S^{2}$. In what follows, the  dyonic black  solution
$\left(\begin{array}{c}
0\\
2
\end{array}\right)$  can be
considered as a four-qubit system supported by string/string duality
in six dimensions \cite{36}.

\section{String/string duality interpretation of four-qubits}
In this section, we  would like to present  a stringy interpretation
of four-qubits using the string/string duality relating type IIA and
heterotic superstings \cite{36}. Indeed,  instead of thinking in
terms of type IIA D-barnes wrapping  non trivial cycles, as done in
the second section of the present work, we consider an equivalent
description in heterotic superstring using cycles in $T^{4}$.  More
precisely, we associate to each element of the cohomology classes of
the heterotic superstring on $T^{4}$ a state of the four-qubit
basis. The basis states can be interpreted in terms of the trivial
fibration $T^{4}=T^{2}\times T^{2}$. To see that, let us consider
the complex realization of $T^{2}\times T^{2}$,
\begin{eqnarray}
z_{\alpha}&=&z_{\alpha}+1\\
z_{\alpha}&=&z_{\alpha}+i, \quad i^2=-1, \quad  \alpha=1,2.\nonumber
\end{eqnarray}
The cohomology classes of this trivial fibration correspond to the
holomorphic and the  anti-holomorphic forms $(p,q)$ which are listed
in table 1.
\begin{table}
\hspace*{\fill}%
\begin{tabular}{|c|c|}
\hline (p,q)forms  & number\tabularnewline \hline \hline 1  &
1\tabularnewline \hline
$dz_{1},\,\overline{dz_{1}},\,dz_{2},\,\overline{dz_{2}}$  &
4\tabularnewline \hline
$dz_{1}\wedge\overline{dz_{1}},\,dz_{1}\wedge
dz_{2},\,dz_{1}\wedge\overline{dz}_{1}$  & 3\tabularnewline \hline
$dz_{2}\wedge\overline{dz_{2}},dz_{2}\wedge
dz_{1},\,dz_{2}\wedge\overline{dz}_{1}$  & 3\tabularnewline \hline
$dz_{1}\wedge\overline{dz}_{1}\wedge\overline{dz}_{2},\,dz_{1}\wedge\overline{dz}_{1}\wedge
dz_{2},\,\overline{dz_{1}}\wedge
dz_{2}\wedge\overline{dz}_{2},\,dz_{1}\wedge
dz_{2}\wedge\overline{dz}_{2}$  & 4\tabularnewline \hline
$dz_{1}\wedge\overline{dz}_{1}\wedge dz_{2}\wedge\overline{dz}_{2}$
& 1\tabularnewline \hline
\end{tabular}\hspace*{\fill}

\protect\protect\protect\protect\protect\caption{$(p,q)$-forms on
$T^2\times T^2$. }
\end{table}

The table  arrangement is motivated from the
$2^{4}=C_{4}^{0}+C_{4}^{1}+C_{4}^{2}+C_{4}^{3}+C_{4}^{4}$ which can
be explored to divide the 2-forms into two categories
\[
\begin{array}{c}
dz_{1}\wedge\overline{dz}_{1},\,\,dz_{1}\wedge dz_{2},\,\,dz_{1}\wedge\overline{dz}_{2}\\
{dz_{2}}\wedge\overline{dz_{2}},\,\,\overline{dz}_{1}\wedge
dz_{2},\,\,\overline{dz}_{1}\wedge\overline{dz_{2}}
\end{array}
\]
as required by the normalized volume form on $T^{2}\times T^{2}$
\begin{equation}
\intop_{T^{4}}dz_{1}\wedge\overline{dz}_{1}\wedge
dz_{1}\wedge\overline{dz}_{2}=1.
\end{equation}

To make contact with four-qubit states, we  consider the following
map applied first  on one factor $T^{2}$
\begin{eqnarray}
\omega_{ij}^{1}=(dz_{1})^{i}\wedge(\overline{dz_{1}})^{j}\,\,\longrightarrow\,\,|ij>\,\,\,i,j=0,1
\end{eqnarray}
producing  the  two-qubit states. Similarly, the basis states of
four-qubits can be obtained by fibering trivially the $T^{2}\times
T^{2}$ complex manifold. Indeed, we define    the  factorization
\begin{equation}
\omega_{ijk\ell}=\omega_{ij}^{1}\wedge\omega_{k\ell}^{2}=(dz_{1})^{i}\wedge(\overline{dz_{1}})^{j}\wedge(dz_{2})^{k}\wedge(\overline{dz_{2}})^{\ell},
\qquad i,j,k,\ell=0,1
\end{equation}
representing the  basis states  of  the four-qubits
\begin{equation}
\omega_{ijkl}\longrightarrow|ijk\ell>.
\end{equation}
The normalized condition may be assured by
\begin{equation}
(\omega_{ijk\ell},\,\omega_{i'j'k'\ell'})=\delta_i^{i'}\delta_j^{j'}\delta_k^{k'}\delta_{\ell}^{\ell'}
\end{equation}
where  the  scalar product can be defined by
\begin{equation}
(\omega_{iji^{'}j^{'}},\,\omega_{klmn})=\int_{T^{2}}\omega_{iji^{'}j^{'}}\wedge\ast\omega_{klmn}.
\end{equation}
Here  $\ast$ is the Hodge duality. In  order to establish a
connection with the dyonic solutions in type IIA supersstring, each
state $|ijk\ell>$ should correspond  to a  charge  of the following
D-brane system
$$
\left\{ D0,\,D2,\,D4,\,D6\right\}
$$
in the presence of $U(1)^{8}$ gauge fields rotated by the
$SO(4)\times SO(4)$ isotropy symmetry. In the  heterotic superstring
side, these vectors, obtained from  the graviton and the
antisymmetric B-field, can be split as
\[
8=1+3+4.
\]
 This
decomposition can be supported by the fact that a real vector of
$SO(4)$ splits under the $\frac{1}{2}\times\frac{1}{2}$ spin
representation of $SU(2)$ as a  1-singlet and  a triplet as follows
\[
4=1+3.
\]
This matches perfectly with the above table of the differential
complex  forms arrangement on the trivial fibration of $T^{2}\times
T^{2}$. In this way, the eight electric charges are linked with
$1+3$ vectors of type $g_{\mu a}$ and 4 vectors of type $B_{\mu a}$
of the heterotic superstring in six dimensions. The 8 magnetic
charges $q_{a}$ can be associated with the dual objects as required
by  the electric and magnetic duality
\begin{equation}
p_{a}q^{a}=2\pi k.
\end{equation}
These objects  forming  a dyonic pair of a black solution
$\left(\begin{array}{c}
0\\
2
\end{array}\right)$ carrying 8 electric and 8 magnetic charges are
associated with the   four-qubit states.

\section{ Quaternionic description of four-qubits}
The quaternionic  character of the moduli space    of type IIA
superstring  on K3 surface, or heterotic superstring on $T^4$ pushes
us to think about a quaternionic  analysis  of four-qubits. This
could help to clarify ceratin issues by drawing a clear contrast
between the attractor mechanism  of black holes developed in
superstring theory and quantum information theory. In particular, we
would like to give such a description using the symmetric space
$\frac{SO(4,4)}{SO(4)\times SO(4)}$ parameterized by 16 scalar
fields associated with a dyonic solution $\left(\begin{array}{c}
0\\
2
\end{array}\right)$ with $8$ electric charges and $8$ magnetic charges. It is recalled that a quaternionic field takes the form
\begin{equation}
q=x_{0}+ix_{1}+jx_{2}+kx_{3},
\end{equation}
where $x_{0},x_{1},x_{2}$ and $x_{3}$ are real numbers, $i,j$ and
$k$ are imaginary numbers  such that
\begin{equation}
\begin{cases}
i^{2}=j^{2}=k^{2}=-1\\
ij=-jk=k,\,jk=-kj=1 & ki=-ki=j.
\end{cases}
\end{equation}
Usually, it is convenient to use the matrix representation of
quaternionic fields. It is defined by
\begin{equation}
q=x_{0}\sigma_{0}+i\vec{x}\vec{\sigma},
\end{equation}
where $\vec{x}=(x_{1},x_{2},x_{3})$ and
$\overrightarrow{\sigma}=(\sigma_{1},-\sigma_{2},\sigma_{3})$ are
the usual Pauli matrices and $\sigma_{0}$ is the $2\times2$ identity
matrix. In this way,  a quaternion number is given by
\begin{equation}
x_{0}+x_{1}i+x_{2}j+x_{3}ij\longrightarrow\left(\begin{array}{cc}
x_{0}+x_{3} & -x_{1}+x_{2}\\
x_{1}+x_{2} & x_{0}-x_{3}
\end{array}\right).
\end{equation}
It turns out that  the scalars of the   dyonic solutions, studied in
the present work,  can be combined to form a quaternionic geometry
in terms of four quaternionic blocks. To see that, we first recall
that these
 scalar fields
belong to the $(4,4)$ bifundamental representation of $SO(4)\times
SO(4)$ symmetry
\begin{equation}
16=(4,4).
\end{equation}
In this way,   they   are specified by two indices $a$ and $b$,
\begin{equation}
\phi\equiv\phi_{b}^{a}.
\end{equation}
Then, we  consider the following decomposition of $SO(4)\times
SO(4)$ symmetry
\begin{equation}
SO(4)\times SO(4)\longrightarrow SU(2)\times SU(2)\times SU(2)\times
SU(2).
\end{equation}
The corresponding representations are given by four integers
$(m_{1},m_{2},m_{3},m_{4})$, where $m_{s}$ are dimensions of
particle state vector spaces. It is recalled that
\begin{equation}
m_{s}=2j_{s}+1,\qquad s=1,2,3,4
\end{equation}
where $j_{s}$ are spin particles. A priori, there are many ways to
decompose  the bifundamental representation $(4,4)$ in terms of
$(m_{1},m_{2},m_{3},m_{4})$. A way, which could be related to
quaternionic geometry, is
\begin{equation}
(4,4)=(4,1,1,1)\oplus(1,4,1,1)\oplus(1,1,4,1)\oplus(1,1,1,4).
\end{equation}
This decomposition shows that the symmetric space
$\frac{SO(4,4)}{SO(4)\times SO(4)}$   can be  parameterized in terms
of   four quaternionic fields associated with 16 charges of the
$\left(\begin{array}{c}
0\\
2
\end{array}\right)$  dyonic black object. In fact,  the scalars  can be combined to form four
quaternionic fields  indicated by only one index
\begin{equation}
\phi_{b}^{a}\longrightarrow\phi_{b_{1}b_{2}}^{a}\longrightarrow\phi_{b_{1}b_{2}}^{a_{1}a_{2}}\longrightarrow\phi^{A},
\quad A=1,2,3,4
\end{equation}
where $a_{1}$ and $a_{2}$ refers to the  $SU(2)$ group, the same for
$b_{1}$and $b_{2}$.   These quaternionic fields can be explored   to
produce a  superpotential of  $N=4$ sigma model in two dimensions
\begin{equation}
W=W(\phi^{1},\phi^{2},\phi^{3},\phi^{4},p^{a},q^{a})\qquad
a=1,\ldots,8.
\end{equation}
 In what follows, we will show that  this superpotential can  be
viewed as a general state of four-qubit systems.  Indeed,  the non
commutativity character of the quaternionic fields can be used to
make contact with a particular class of the  Clifford algebras. In
this way, the superpotentiel $W(\phi)$ can be interpreted as an
element of such   a  Clifford algebra.   Assuming that  the fields
$\phi_A$ form a normalized basis of  a  vector space $V$ and using
the work  developed in \cite{27},  the  algebra spanned by the all
reduced products of the form
\begin{equation}
Span\left\{
\phi_{1}^{i}\phi_{2}^{j}\phi_{3}^{k}\phi_{4}^{l}\right\}, \qquad
i,j,k,\ell=0,1
\end{equation}
defines a class of  the Clifford algebras,  denoted as $Cl_{0,4}$.
It has been shown   that  this  algebra could be  decomposed as
follows
\begin{equation} Cl_{0,4}={\oplus}_{k=0}^4Cl_{0,4}^{(k)}
\end{equation}
where  $Cl_{0,4}^{(k)}$ is known by the space of  $k$-multivectors
\cite{27}.
 In connection with the  heterotic  superstring
compactification,  a close inspection shows that the algebra
$Cl_{0,4}$ can be associated with the cohomology classes of $T^{4}$.
More precisely, we have
\begin{equation}
\left(\begin{array}{c}
Cl_{0,4}^{0}\,\,:\,\mbox{the scalar}\\
Cl_{0,4}^{1}\,\,:\,\,\mbox{the 1-forms}\\
Cl_{0,4}^{2}\,\,:\,\,\mbox{the 2-forms}\\
Cl_{0,4}^{3}\,\,:\,\,\mbox{the 3-forms}\\
Cl_{0,4}^{4}\,\,:\,\, \mbox{the volume form}
\end{array}\right).
\end{equation}
Inspired by such a decomposition, we  propose  the  mapping
\[
|ijkl>\longrightarrow\phi_{1}^{i}\phi_{2}^{j}\phi_{3}^{k}\phi_{4}^{l},
\qquad i,j,k,\ell=0,1.
\]
In this way, the  general state of four-qubits corresponds to  a
quaternionic  superpotential
\begin{equation}
|\psi\rangle=\sum\limits _{ijk\ell=0,1}a_{ijk\ell}|ijk\ell\rangle
\longrightarrow
W=W(\phi_{1},\phi_{2},\phi_{3},\phi_{4},p^{a},q^{a}).
\end{equation}
In this mapping, the  $a_{ijk\ell}$ numbers should correspond to the
$(p^a,q^a)$ charges of  the dyonic object $\left(\begin{array}{c}
0\\
2
\end{array}\right)$ .   The  superpotential could  be
viewed as the holomorphic sections of the line bundles on four
dimensional quernionic manifolds. We expect that these sections may
be encoded in  non trivial polytopes going beyond the toric graphs
associated with the projective complex geometry used in \cite{37}.

\section{Conclusion and open questions}

In this work, we have approached  four-qubit systems  in the context
of type II superstring compactifications using string dualities
between type IIA and heterotic superstrings. This connection has
been elaborated by giving a classification according to black brane
dimensions, which results in two kinds of dyonic solutions
generalized in arbitrary dimension where the electric and magnetic
charges have been linked formally to Calabi-Yau Hodge numbers. We
have shown that the four-qubit systems are related to a stringy
moduli space $\frac{SO(4,4)}{SO(4)\times SO(4)}$, which is a
reduction of a moduli space of the  heterotic superstring $T^4$
living in six dimentional supergravity model.  Using string/string
duality, the four-qubit states are related to  a dyonic solution
BH-B2B
 $\left(\begin{array}{c}
0\\
2
\end{array}\right)$ carrying 8 electric and 8 magnetic charges $
(p_{a},q_{a})$. Moreover, it has been remarked that the usual
decomposition of $SO(4)\times SO(4)\longrightarrow SU(2)\times
SU(2)\times SU(2)\times SU(2)$ results in four quaternionic fields
that can be interpreted as  states of  the four-qubits. These states
are linked with a  quaternionic superpotentiel $W(\phi)$ of $N=4$
sigma interpreted as an element of a particular class of the
Clifford algebras  denoted as  $Cl_{0,4}$. \\
 The present work comes up with  certain
open questions related to quantum information theory.  It is
recalled  that interesting  works dealing with four-qubits from
algebraic geometry point of  view including ADE singularities have
been elaborated in \cite{38,39}.  It would be interesting to see if
this has any possible connection with such activities.   Moreover,
it should be of relevance to approach quantum information concepts
using quaternionic geometry associated with Clifford algebras. This
includes the study of entanglement and quantum discord. It is
clearly interesting to better understand such concepts from
geometric methods.  We anticipate that many concepts used in quantum
information, of four-qubits,  could be discussed using quaternionic
manifolds. This
 will be addressed
elsewhere. \\
\\
{\bf Acknowledgments}:   The authors would like to thank M. Asorey
for discussions. AB would like to thank the Departamento de
F\'{\i}sica T\'{e}orica, Universidad de Zaragoza  for very kind
hospitality and scientific  supports during the realization of a
part of this work. He also acknowledges the warm hospitality of
 Montanez and  Naz  families durante his travel in Spain
and he thanks also Hajja Fatima (his mother) for patience and
supports. AS is supported by FPA2012-35453.


\begin{thebibliography}{99}

 \bibitem{1}

 A. Strominger, C. Vafa, {\em Microscopic Origin of the Bekenstein-Hawking Entropy}, Phys.Lett.
{\bf B379} (1996) 99, {\tt arXiv:hep-th/9601029}.

 \bibitem{2}

 C. Vafa,  {\em Black Holes and Calabi-Yau Threefolds}, Adv.Theor.Math.Phys. {\bf 2} (1998)
 207, {\tt
hep-th/9711067}.

 \bibitem{3}

 J. Maldacena, A. Strominger, E. Witten, {\em Black Hole Entropy in M-Theory}, JHEP{\bf 9712} (1997)002, {\tt  arXiv:hep-th/9711053}.

 \bibitem{4}

  B. Haghighat, S. Murthy, C. Vafa, S. Vandoren, {\em  F-Theory, Spinning Black Holes and
Multistring Branches},   {\tt arXiv:1509.00455}.

 \bibitem{5}

  S. Ferrara, R. Kallosh, A. Strominger,  {\em N = 2 Extremal Black Holes}, Phys. Rev.  {\bf
  D52}
(1995) 5412, {\tt hep-th/9508072}.

 \bibitem{6}

 S. Ferrara and R. Kallosh,  {\em Supersymmetry and Attractors}, Phys. Rev. {\bf D54} (1996) 1514,
{\tt hep-th/9602136}.
 \bibitem{7}

 R. Ahl Laamara, M. Asorey, A. Belhaj, A, Segui, {\em  Extremal Black Brane Attractors on The
Elliptic Curve}, J.Phys. {\bf A43} (2010) 105401, {\tt
arXiv:0907.0093}.

 \bibitem{8}

 P. Bueno, R. Davies, C. S. Shahbazi,  {\em Quantum black holes in Type-IIA String
 Theory}, {\tt
arXiv:1210.2817}.

 \bibitem{9}

H. Ooguri, A. Strominger, C. Vafa, {\em Black Hole Attractors and
the Topological String}, Phys.Rev. {\bf D70}(2004)106007, {\tt
arXiv:hep-th/0405146}.

 \bibitem{10}

 S. Bellucci, S. Ferrara, A. Marrani and A. Yeranyan,  {\em Mirror Fermat Calabi-Yau threefolds
and Landau-Ginzburg Black Hole Attractors}, Riv. Nuov o Cim. {\bf
029}  (2006)1, {\tt hep-th/0608091}.


 \bibitem{11}

 A. Belhaj, {\em On Black Objects in Type IIA Superstring Theory on Calabi-Yau Manifolds},
African Journal Of Math. Phys. Vol. {\bf 6} (2008)49, {\tt
arXiv:0809.1114}.

 \bibitem{12}

 M. J. Duff, S. Ferrara, A. Marrani,{\em
D = 3 Unification of Curious Supergravities}, JHEP {\bf 1701} (2017)
023, {\tt arXiv:1610.08800 [hep-th]}.




 \bibitem{13}

 A. Belhaj, M. Bensed, Z. Benslimane, M. B. Sedra, A. Segui, {\em Qubit and Fermionic Fock
Spaces from Type II Superstring Black Hole},   Int. J. Geom. Methods
Mod. Phys. {\bf 14}  (2017)1750087 {\tt arXiv:1604.03998}.



 \bibitem{14}

  A. Belhaj, Z. Benslimane, M. B. Sedra, A. Segui, {\em Qubits from Black Holes in M-theory on
K3 Surface},   Int. J. Geom. Methods Mod. Phys. {\bf 13}
(2016)1650075 {\tt arXiv:1601.07610}.

 \bibitem{15}
 M. J. Duff, S. Ferrara, {\em Four curious supergravities}, Phys. Rev. {\bf D83} (2011)046007, {\tt arXiv:1010.3173}.


 \bibitem{16}
 P. Levay,  {\em Qubits from extra dimensions}, Phys. Rev. {\bf D84} (2001)125020.

 \bibitem{17}

 M. J. Duff, String triality, {\em  black hole entropy and Cayley s hyperdeterminant}, Phys. Rev.
{\bf D76} (2007) 025017, {\tt hep-th/0601134}.

 \bibitem{18}

 P. Levay, {\em Stringy Black Holes and the Geometry of Entanglement }, Phys. Rev.  {\bf D74}, 024030 (2006), {\tt  arXiv:0603136}.

\bibitem{19}
 P. Levay, F. Holweck, {\em  Embedding qubits into fermionic Fock space, peculiarities of the four-qubit
case}, (2015), {\tt  arXiv:1502.04537}.



 \bibitem{20} P. Levay,
F.  Holweck, M. Saniga,  {\em The magic three-qubit Veldkamp line: A
finite geometric underpinning for form theories of gravity and black
hole entropy}, {\tt
  arXiv:1704.01598}.


  \bibitem{21}
 M. Cvetic, G.W. Gibbons, C.N. Pope, {\em
Compactifications of Deformed Conifolds, Branes and the Geometry of
Qubits},  {\tt arXiv:1507.07585}.





 \bibitem{22}
L. Borsten, D. Dahanayake, M. J. Duff, A. Marrani, W. Rubens, {\em
Four-qubit entanglement from string theory}, Phys.Rev.Lett. {\bf
105} (2010)100507,  {\tt arXiv:1005.4915}.


\bibitem{23}
 L. BorstenM. J. DuffA. Marrani,W. Rubens, {\em
On the Black-Hole/Qubit Correspondence}, Eur. Phys. J. Plus {\bf
126} (2011)37 , {\tt [arXiv:1101.3559 [hep-th]}.




 \bibitem{24}

 Y. Aadel, A. Belhaj, M. Bensed, Z. Benslimane, M. B. Sedra, A. Segui, {\em  Qubit Systems from Colored Toric Geometry and Hypercube Graph Theory},  .
Commun. Theor. Phys. {\bf 68}(2017) 285

 \bibitem{25}

 A. Belhaj, M. B. Sedra, A. Segui,  {\em Graph Theory and Qubit Information Systems of
Extremal Black Branes},  J.Phys. {\bf A48} (2015)045401, {\tt
arXiv:1406.2578}.

 \bibitem{250}
 A. Belhaj, {\em  Multi-qubits and Polyvalent
Singularity in Type II Supestring Theory}, {\tt  arXiv:1612.09356}.

 \bibitem{251}
A. Belhaj, A. Belhaj, L. Machkouri, M. M. Sedra, S. Ziti, {\em Graph
Theory Representation of Quantum Information Inspired by Lie
Algebras}, {\tt  arXiv:1609.03534}.

\bibitem{26}

P. Levay, {\em  STU Black Holes as Four Qubit Systems}, Phys. Rev.
{\b D82}(2010)026003, {\tt arXiv:1004.3639}.








\bibitem{260}
P. Levay, M. Planat, Metod Saniga,  {\em  Grassmannian Connection
Between Three- and Four-Qubit Observables, Mermin's Contextuality
and Black Holes},  JHEP {\bf 09} (2013)037, {\tt arXiv:1305.5689}.



 \bibitem{27}
J. A. Emanuello, {\em Analysis of Functions of Split-Complex,
Multicomplex, and Split-Quaternionic Variables and Their Associated
Conformal Geometries}. PhD thesis, The Florida State University,
2015.
\bibitem{270}
 A. Belhaj, {\em  Manifolds of $G_2$ Holonomy
from N=4 Sigma Model}, J.Phys. {\bf A35}(2002)8903, {\tt
arXiv:hep-th/0201155}.
\bibitem{28}
P. Candelas, G. Horowitz, A. Strominger, E. Witten,  {\em Vacuum
configurations for superstrings}, Nucl. Phys.  {\bf B258} (1985)46.

 \bibitem{29}
B.R. Greene,  {\em String Theory on Calabi Yau Manifolds},  {\tt
hep-th/9702155}.

 \bibitem{30}
 P. Aspinwall, {\em K3 surfaces and String Duality},  {\tt hep-th/961117}.

\bibitem{31}
A. Belhaj, L.B. Drissi, E.H. Saidi, A. Segui,  {\em N=2
Supersymmetric Black Attractors in Six and Seven Dimensions}, Nucl.
Phys. {\bf B796} (2008)521, {\tt arXiv:0709.0398}.

\bibitem{310}
 E.H. Saidi, A. Segui, {\em Entropy of Pairs of Dual
Attractors in 6D/7D},  JHEP {\bf 0807}(2008)128, {\tt
arXiv:0803.2945}.

\bibitem{32}
G. Ottavian,  {\em Introduction to the Hyperdeterminant and to the
Rank of Multidimensional Matrices},  {\tt arXiv:1301.0472}.

\bibitem{320}
 A. Cayley, {\em
On the theory of linear transformations}, Camb. Math. J. 4
193-209,1845.



\bibitem{321}
 I.M. Gelfand, M.M. Kapranov, A.V. Zelevinsky {\em Discriminants, Resultants and
Multidimensional Determinants}, Birkhauser, 1994.


\bibitem{33}
M. A. Nielsen, I. L. Chuang, {\em Quantum Computation and Quantum
Information}, Cambridge University Press, New York, NY, USA, 2000.

\bibitem{34}
D. R. Terno, {\em Introduction to relativistic quantum information},
{\tt  arXiv:quant-ph/0508049}.


\bibitem{35}
M. Kargarian,  {\em Entanglement properties of topological color
codes}, Phys. Rev. {\bf A78} (2008)062312, {\tt arXiv:0809.4276}.


\bibitem{36}
 C.
Vafa, {\em Lectures on Strings and Dualities}, {\tt
arXiv:hep-th/970220}.

\bibitem{37}
A. Belhaj, H. Ez-Zahraouy, M. B. Sedra,  {\em Toric Geometry and
String Theory Descriptions of Qudit Systems}, J. Geom. Phys. {\bf
95} (2015)21, {\tt  arXiv:1408.3952}.
\bibitem{38}
F. Holweck, J-G. Luque, M. Planat,  {\em Singularity of type $D_4$
arising from four qubit systems},  {\tt arXiv:1312.0639}.
\bibitem{39}
 F. Holweck, H. Jaffali, {\em
Three-qutrit entanglement and simple singularities}, {\tt
arXiv:1606.05537}.

\bibitem{40}
  R.Kallosh, A.D.Linde, {\em Strings, black holes, and quantum information},Phys.Rev. D {\bf 73} (2006) 104033 {\tt
arXiv:hep-th/0602061}.

\bibitem{41}
S.Ferrara, A.Marrani, {\em Symmetric Spaces in Supergravity},
Contemp. Math. {\bf 490}(2009) 203, {\tt arXiv:0808.3567 [hep-th]}.


\end{thebibliography}
\end{document}